\documentstyle[11pt,psfig]{article}
\begin{document}
\title{\huge{$J/\psi$ Production at the LHC}
\thanks{Research partially supported by CICYT under grant AEN-96/1718}} 
\author{{\bf B. Cano-Coloma\thanks{bcano@evalo1.ific.uv.es} $\ $ and M.A. 
Sanchis-Lozano\thanks{mas@evalvx.ific.uv.es}}
\\ \\
\it Instituto de F\'{\i}sica
 Corpuscular (IFIC), Centro Mixto Universidad de Valencia-CSIC \\
\it and \\
\it Departamento de F\'{\i}sica Te\'orica \\
\it Dr. Moliner 50, E-46100 Burjassot, Valencia (Spain) }
\maketitle 
\begin{abstract}
  We firstly re-examine  hadroproduction of  prompt $J/\psi$'s at the 
  Fermilab Tevatron finding that  those colour-octet   matrix  elements 
  presented in literature  are systematically overestimated  due to the 
  overlooking of   the effective primordial   transverse  momentum 
  of  partons  (i.e.   dynamically generated via initial-state radiation). 
  We estimate the  size of these effects using   different parton  
  distribution functions in a Monte Carlo framework.  Finally, we extrapolate 
  up to LHC energies making a prediction on the expected $p_t$ differential 
  cross-section for charmonium. 
\end{abstract}
\vspace{-14.5cm}
\large{
\begin{flushright}
  IFIC/97-1\\
  FTUV/97-1\\
  \today
\end{flushright} }
\vspace{15.5cm}
{\small PACS numbers: 12.38.Aw; 13.85.Ni} \\
{\small keywords: Quarkonia production; Colour-Octet; NRQCD; LHC; Tevatron }
\newpage
\section{Introduction}
Heavy quarkonia have been playing an important role over the past decades
in the development of the theory of the strong interaction and
such beneficial influence seems far from ending. Indeed, the experimental
discovery at Fermilab \cite{fermi} of an excess of inclusive
production  of prompt heavy  quarkonia (mainly  for  $J/\psi$ and  $\psi(2S)$
resonances) in  antiproton-proton collisions triggered an intense theoretical
activity beyond what was considered  conventional  wisdom until recently
\cite{greco,baranov}. The  discrepancy between the  so-called colour-singlet
model (CSM) in hadroproduction \cite{baier} and the experimental data amounts
to more  than an order   of magnitude  and   cannot  be attributed to   those
theoretical uncertainties  arising from the  ambiguities on  the choice of  a
particular    parton distribution function (PDF),    the  heavy quark mass or
different energy \vspace{0.1in} scales.
\par 

Recently it  has been argued that the  surplus of charmonia production can be
accounted for by assuming that the heavy quark pair not necessarily has to be
produced in a colour-singlet  state  at the short-distance partonic   process
itself \cite{braaten}.   Alternatively, it can  be  produced in a  colour-octet
state evolving non-perturbatively into  quarkonium in a specific  final state
with the correct quantum  numbers according to some computable  probabilities
governed by the internal velocity of the heavy quark. This mechanism, usually
named as   the colour-octet  model  (COM),   can be  cast into  the  rigorous
framework of an effective non-relativistic theory for the strong interactions
(NRQCD) deriving from first principles \vspace{0.1in} \cite{bodwin}.
\par

However, the weakness  of the COM lies  in the fact that the non-perturbative
parameters characterizing  the long-distance hadronization process beyond the
colour-singlet contribution  (i.e. the colour-octet  matrix elements)  are so
far almost free parameters to be adjusted  from the fit to experimental data,
though expected to be  mutually   consistent according  to the  NRQCD   power
counting \vspace{0.1in} rules.
\par 

On  the other  hand, an  attractive   feature of the colour-octet  hypothesis
consists of the  universality of the  NRQCD matrix elements entering in other
charmonium production processes like photoproduction from electron-proton
collisions at HERA \cite{cacciari}.  Let us look below in some detail at the
way hadronization is folded with the partonic  description of hadrons, focusing
for concreteness on the couple of related papers \vspace{0.1in} 
\cite{cho0,cho}. 
\par

In the first paper \cite{cho0},  Cho and Leibovich considered for  charmonium
production (in addition to the CSM) the $^3S_1$ coloured intermediate state as 
a first approach, computing the squared amplitudes as products of perturbative
parts standing for the short-distance partonic processes, and the colour-octet 
matrix element concerning the long-distance hadronization. Finally, a
convolution of concrete  parton distribution functions and the differential 
cross-section for the $Q\bar{Q}$  production subprocess was performed, whereby
the $p_t$ dependence of the  charmonia production exclusively coming from 
the \vspace{0.1in} latter.
\par 
In Ref. \cite{cho} the same
authors take into account further contributions from new coloured states (for
more details see  the quoted references) concluding finally   that at high
enough $p_t$ a two-parameter fit is actually required to explain the observed
inclusive  $p_t$   distribution  of charmonia   production  at  the 
\vspace{0.1in} Tevatron.
\par

Notice, however, that such calculations were carried out based on a possibly
oversimplified picture of the hadronic interaction. Indeed, it is  
already well-known a long time ago that higher-order QCD effects  ($K$
factors)  play an important  role  in  inclusive  hadroproduction. In 
particular, beyond the primordial transverse  momentum $k_t$ of
partons in hadrons due to Fermi motion relevant at small $p_t$,
initial-state radiation of gluons  by the interacting partons add up
yielding an {\em effective} intrinsic transverse momentum which certainly has 
to  be considered in hadroproduction at high $p_t$ \cite{break}. As we 
shall see, if overlooked at all in charmonia production, the specific effect
on the fit parameters  (and ultimately on the colour-octet
matrix elements) amounts to a {\em systematic} \vspace{0.1in} overestimate. 

\section{Extraction of NRQCD matrix elements from Tevatron data}

In   this  work we  have  implemented the colour-octet mechanism 
in the  event   generator PYTHIA 5.7 \cite{pythia}  via the following
${\alpha}_s^3$ partonic processes: $g+g{\rightarrow}J/\psi+g$, 
$q+g{\rightarrow}J/\psi+q$ and $q+\bar{q}{\rightarrow}J/\psi+g$ 
\footnote{Originally  PYTHIA generates   direct
production of $J/\psi$'s via the CSM \cite{pythia}. In an earlier work
\cite{bea} we considered only the gluon-gluon fusion but with the
colour mechanism implemented in}, including the $^3S_1$ and
$^1S_0+^3P_0$ contributions as coloured intermediate states. The
corresponding differential cross sections can be found in \cite{cho0,cho}. We 
conclude from our simulation that gluon-gluon fusion actually stands for 
the dominant process as expected, gluon-quark
scattering contributes appreciably, whereas 
quark-antiquark scattering represents a tiny contribution. 
Let us stress that initial- and final-state
radiation were incorporated  within the  Lund Monte  Carlo
framework \cite{sjos}. Of course, the PYTHIA treatment of the effective 
$k_t$ is not guaranteed to be perfect but, nevertheless, should  give a 
reasonable estimate of such \vspace{0.1in} effects.
\par

\begin{figure}[htb]
\centerline{\psfig{figure=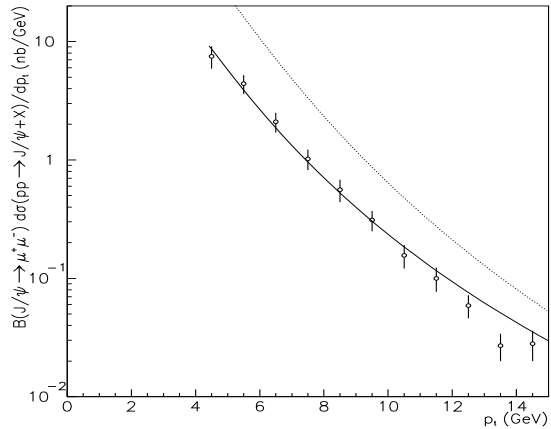,height=6.5cm,width=8.cm}}
\caption{ Curves obtained from PYTHIA (not fit) including the colour-octet
mechanism for prompt $J/\psi$ production at the Tevatron using the same
parameters as in Ref. [9]. The solid line corresponds to initial- and 
final-state radiation off and the dotted line to  initial- and
final-state radiation on. The MRSD0 parton distribution
function and $M_c=1.48$ GeV were employed as in [9].} 
\end{figure}

Using  the  same numerical  values  for the  colour-octet matrix  elements as
reported in tables I and II of Ref. \cite{cho}, if initial- and final-state
radiation are  turned off (there exists this possibility in PYTHIA 
\cite{pythia}), there is a good agreement between the theoretical curve and
the experimental points  (see Fig. 1), as should be reasonably 
\vspace{0.1in} expected. 
\par
However, if initial- and final-state radiation \footnote{It
should  be noted that  initial-state radiation and final-state radiation have
opposite  effects in the $p_t$ spectrum,  the former enhancing the high $p_t$
tail  whereas  the latter  softens the  distribution}  are switched on, the
predicted curve stands  well above  the  experimental data over the $p_t$ 
range examined, in accordance with the expected ${\lq}{\lq}$kick" caused by
the {\em effective} primordial transverse momentum of partons \cite{break}. 
Accordingly, keeping radiation effects on in the modelling of the
full hadronic production process, it turns out that the values of the
colour-octet  matrix elements have to be lowered by significant 
\vspace{0.1in} factors.
\par 
\newpage

In  order to assess the impact of the effective intrinsic $k_t$ upon
the NRQCD matrix elements we have made  three different choices for the
proton PDF \footnote{See \cite{pdflib} for  technical  details about   the
package of  Parton Density Functions  available at  the CERN Program Library.
References therein}: 

\begin{description} 
 \item[a)] the leading order CTEQ 2L (by default in PYTHIA 5.7)  
 \item[b)] the next to leading order MRSD0 (the same as used in \cite{cho}) 
 \item[c)] the next to leading order GRV 94 HO 
\end{description}

As already mentioned  before, the  theoretical  curve of the  inclusive $p_t$
distribution of prompt  $J/\psi$'s   stands   in all cases  above    Tevatron
experimental  points if  the set of  parameters from  \cite{cho}  are blindly
employed in the PYTHIA generation with initial-state radiation on.  Motivated
by this systematic discrepancy, we performed new fits for the prompt $J/\psi$
direct   production  at   Tevatron    (feed-down  from  radiative decay    of
${\chi}_{cJ}$  resonances  was   experimentally removed).   The corresponding
colour-octet matrix elements are shown in table 1 and the plots for cases a) 
and b) in Fig. 2. We found ${\chi}^2/N_{DF}=1.8,\ 2.5$ and $1.1$ for
a) , b) and c) \vspace{0.1in} respectively.
\par 
Notice that our new results for the matrix elements are in general slightly
smaller than those presented in our previous work \cite{bea}. This is in
accordance with the fact that we are including new contributions not
considered before so that the fit parameters have to be further lowered 
accordingly.

\begin{figure}[htb]
\centerline{\hbox{
\psfig{figure=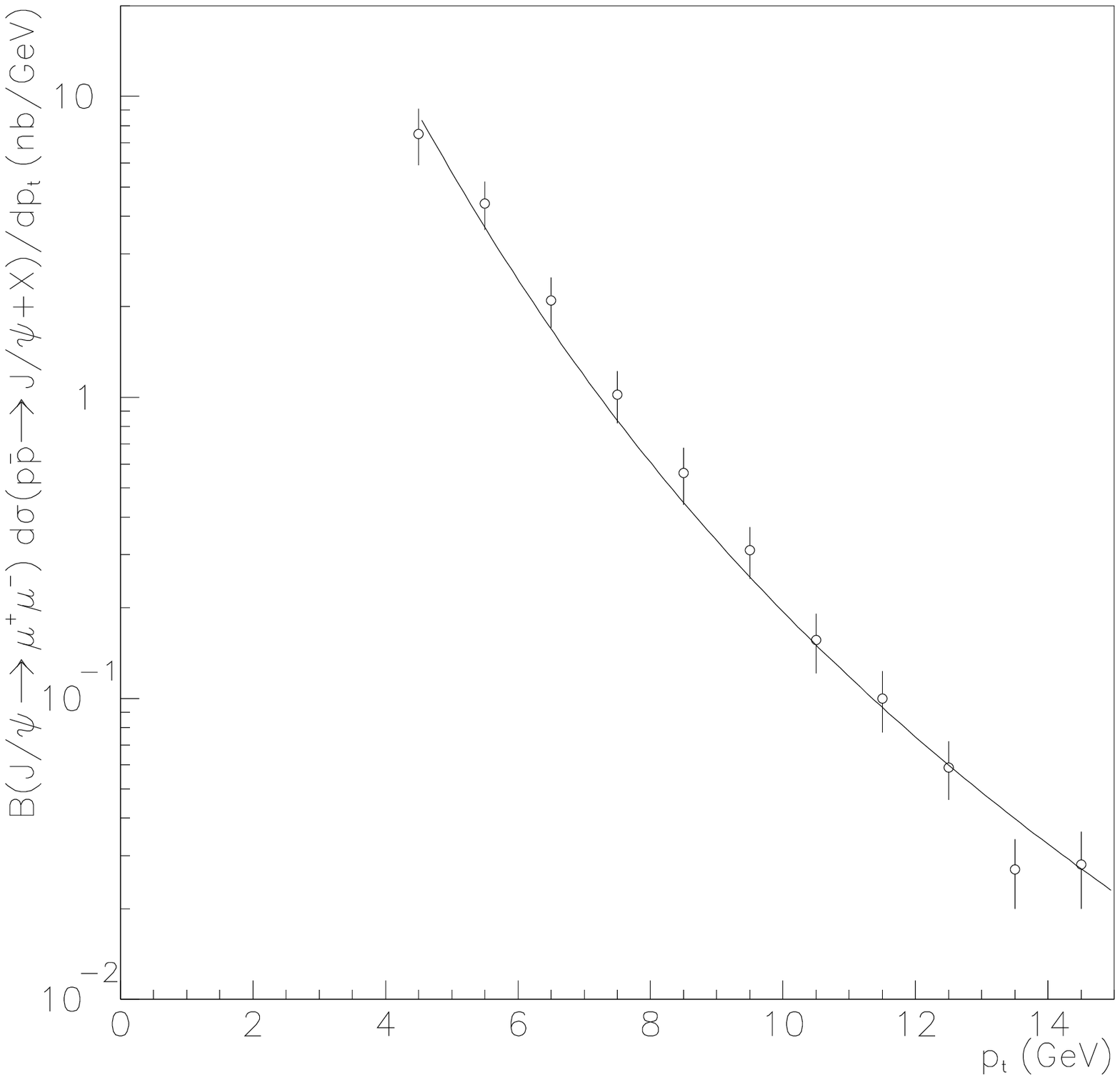,height=6.5cm,width=8cm}
\psfig{figure=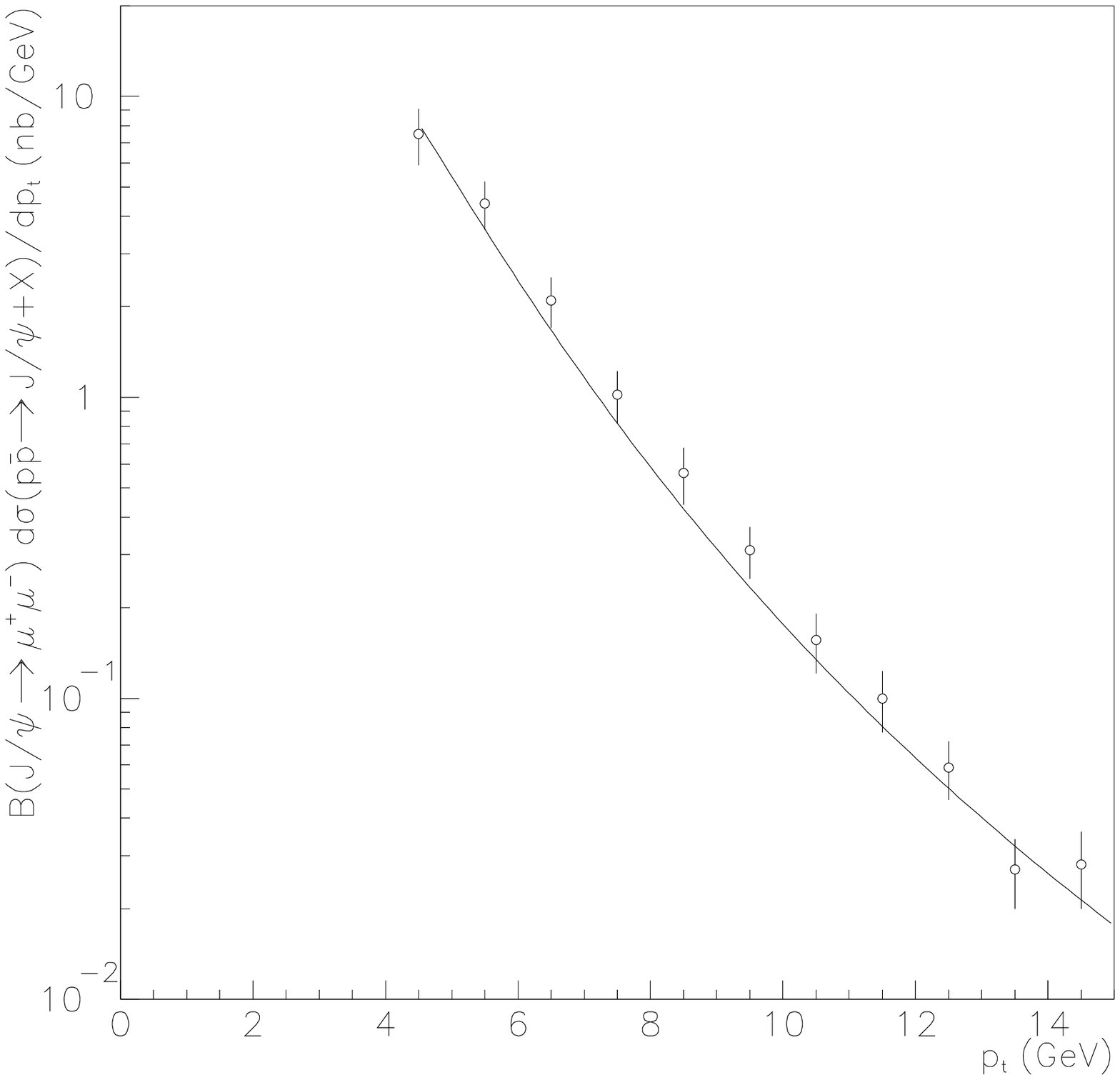,height=6.5cm,width=8cm}
}}
\caption{Two-parameter fits to the experimental Tevatron data using
CSM + COM, where initial- and final-state radiation were incorporated 
via PYTHIA generation. The CTEQ 2L (MRSD0) parton distribution function 
was employed in the left (right) plot.} 
\end{figure}

\begin{table*}[hbt]
\setlength{\tabcolsep}{1.5pc}
\newlength{\digitwidth} \settowidth{\digitwidth}{\rm 0}
\caption{Colour-octet matrix elements (in units of GeV$^3$) from the 
best fit to Tevatron data on prompt $J/\psi$ production
using different parton distribution functions. The error bars are
statistical only. For comparison we quote the values given in Ref. [9]: 
$(6.6{\pm}2.1){\times}10^{-3}$ and $(2.2{\pm}0.5){\times}10^{-2}$  
respectively.}
\label{FACTORES}

\begin{center}
\begin{tabular}{lcc}    \hline
matrix element:  & $<0{\mid}O_8^{J/\psi}(^3S_1){\mid}0>$ & 
$\frac{<0{\mid}O_8^{J/\psi}(^3P_0){\mid}0>}{M_c^2}+
\frac{<0{\mid}O_8^{J/\psi}(^1S_0){\mid}0>}{3}$ \\
\hline

CTEQ2L &$(3.3{\pm}0.5){\times}10^{-3}$& $(4.8{\pm}0.7){\times}10^{-3}$ \\
MRSD0 & $(2.1{\pm}0.5){\times}10^{-3}$ & $(4.4{\pm}0.7){\times}10^{-3}$ \\
GRV 94 HO & $(3.4{\pm}0.4){\times}10^{-3}$ & $(2.0{\pm}0.4){\times}10^{-3}$ \\
\hline
\end{tabular}
\end{center}
\end{table*}

\section{Extrapolation to LHC}

Finally,  we  have  generated  prompt $J/\psi$'s in   proton-proton
collisions at LHC  energies (center-of-mass energy = 14  TeV) by means of our
${\lq}{\lq}$modified" version of PYTHIA with the colour-octet matrix elements
as shown in Table 1 - i.e. after normalization to  Tevatron data. Indeed, an
order-of-magnitude estimate  for the  expected  production of
charmonia at the  LHC is suitable from  many points of  views \cite{atlas}:
$J/\psi$ can be considered either as a signal in its own right or as a source
of background for other  interesting processes involving $J/\psi$'s, like  CP
studies from  the cascade channel \vspace{0.1in} 
$B_d^0\  {\rightarrow}\ J/\psi\ K_s^0$. 

\begin{figure}[htb]
\centerline{\psfig{figure=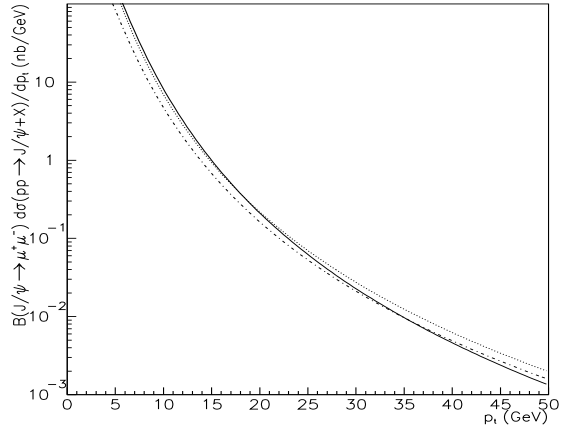,height=6.5cm,width=8.cm}}
\caption{Our prediction for prompt $J/\psi$ 
direct production at the LHC using PYTHIA with the colour-octet matrix
elements from Table 1: a) dotted line: CTEQ 2L, b) dot-dashed line: MRSD0, 
and c) solid line: GRV HO. The rapidity cut 
${\mid}y{\mid}<2.5$ on the $J/\psi$ was required.
} 
\end{figure}

\par
In Fig. 3 we show altogether the $p_t$ inclusive distributions of
direct prompt $J/\psi$'s at the LHC obtained for the three  PDF's employed
in our study. Comparing them with the distribution obtained by Sridhar 
\cite{sridhar}, we find our predictions standing below his theoretical curve
at large $p_t$. In fact this is not surprising since Sridhar considered only
the $^3S_1$ coloured intermediate state, whose NRQCD matrix element was
taken larger than the one used in this work. The argument ends by noticing 
that at high enough $p_t$ the dominant production comes from the $^3S_1$
contribution as the combined $^1S_0+^3P_0$ contribution falls off 
\vspace{0.1in} faster.
\par
\section{Conclusions}
In this paper we have extended our preliminary study \cite{bea} 
considering all three possible short-distance processes, namely
$g+g{\rightarrow}J/\psi+g$, $q+g{\rightarrow}J/\psi+q$ and
$q+\bar{q}{\rightarrow}J/\psi+g$, contributing at order ${\alpha}_s^3$
to charmonia production including the colour-octet \vspace{0.1in} mechanism.
\par
We have investigated  higher-order effects induced by initial-state radiation
on the extraction  of the NRQCD matrix elements  from hadroproduction at high
$p_t$   by  means of   an event   generator  (PYTHIA 5.7)   with colour-octet
mechanisms   implemented in.  We  conclude  that the overlooking of the 
{\em effective}  primordial $k_t$ leads systematically to a
significant  overestimate  of  the  colour parameters.  We have 
re-calculated those colour-octet matrix elements for $J/\psi$ 
production from fits to Tevatron data using
three  different sets of proton \vspace{0.1in} PDF's.
\par
Finally we have estimated the prompt $J/\psi$ production at the LHC, finding
an overall relative good accordance among the three choices for the
proton PDF. Under the assumption of the validity of the COM, our prediction 
can be used for simulation purposes at LHC experiments (see the Web page at 
http://wwwcn.cern.ch/$^{\sim}$msmizans/
production/0.html for a general view of the B physics simulation status
in ATLAS).

\subsection*{Acknowledgments}

We thank S. Baranov, P. Eerola, N. Ellis and M. Smizanska and the ATLAS B
physics working group for useful comments and an encouraging attitude. 
Comments by E. Kovacs, L. Rossi, T. Sj\"{o}strand and K. Sridhar are also 
acknowledged. 

\end{document}